# 'Getting out of the closet': Scientific authorship of literary fiction and knowledge transfer

Joaquín M. Azagra-Caro[1] · Anabel Fernández-Mesa[2] · Nicolás Robinson-García[1]*

**Abstract** Some scientists write literary fiction books in their spare time. If these books contain scientific knowledge, literary fiction becomes a mechanism of knowledge transfer. In this case, we could conceptualize literary fiction as non-formal knowledge transfer. We model knowledge transfer via literary fiction as a function of the type of scientist (academic or non-academic) and his/her scientific field. Academic scientists are those employed in academia and public research organizations whereas non-academic scientists are those with a scientific background employed in other sectors. We also distinguish between direct knowledge transfer (the book includes the scientist's research topics), indirect knowledge transfer (scientific authors talk about their research with cultural agents) and reverse knowledge transfer (cultural agents give scientists ideas for future research). Through mixed-methods research and a sample from Spain, we find that scientific authorship accounts for a considerable percentage of all literary fiction authorship. Academic scientists do not transfer knowledge directly so often as non-academic scientists, but the former engage into indirect and reverse transfer knowledge more often than the latter. Scientists from History stand out in direct knowledge transfer. We draw propositions about the role of the academic logic and scientific field on knowledge transfer via literary fiction. We advance some tentative conclusions regarding the consideration of scientific authorship of literary fiction as a valuable knowledge transfer mechanism.

**JEL codes** O33 • O34 • Z11

---

* ✉ Joaquín M. Azagra-Caro, jazagra@ingenio.upv.es

[1] INGENIO (CSIC-UPV), Universitat Politècnica de València, Camino de Vera s/n, E-46022 Valencia, Spain

[2] University of Valencia, Valencia (Spain)

## 1 Introduction

Academic scientists, understood as members of the teaching and/or research staff of a public university or a public research organization (including scholars from the Humanities and Social Sciences), benefit the academic community, industry and other social collectives through teaching, research and knowledge transfer (Olmos-Peñuela et al. 2014). Typical knowledge transfer mechanisms include patent licensing, spinoffs, joint projects, research contracts, consulting, informal meetings, etc. (Bercovitz and Feldman 2006; Link et al. 2007; Bradley et al. 2013). Other mechanisms have not deserved so much attention, maybe because they are related to unusual academic activities: participation in fora like diplomatic circles (Fähnrich 2015), political parties (Parker 2015) or engagement with cultural audiences (Benneworth 2014). We address this last aspect, namely sociocultural contributions of academics, given the many voices advocating for its relevance. The role of universities as anchor institutions (Ehlenz 2015), and of scientists as part of a creative class, is important for the sociocultural life of cities (Florida 2005), and for the inclusion of the sociocultural benefits of research in the evaluation of its impact (Goddard 2009). In this paper, we will focus on the conceptualization of one sociocultural contribution of academics as a knowledge transfer mechanism: writing literary fiction. There are many examples of scientists that have written literary fiction. Eminent cases are Nobel Prize Winners like José Echegaray (1904, Technical University of Madrid), Toni Morrison (1993, Texas Southern University and Howard University), and John Maxwell Coetzee (2003, State University of New York and University of Chicago). Some extremely popular authors are also academics, like J.R.R. Tolkien (Oxford University) and Umberto Eco (University of Bologna).[1]

Of course, academic scientists' literary work may not relate to their research. When there is a relation, literary fiction becomes a dissemination channel and, as such, opens an array of policy tools to increase the transfer of knowledge and promote public understanding of science. And even when this fiction is not related to their research, it can still become a mechanism to interact with non-academic stakeholders. By analyzing knowledge transfer via literary fiction, we call attention to problems like employee burnout because of lack of institutional support to creativity in scientific organizations; brain drain of individuals with useful writing skills who leave academia once their literary work becomes successful; or valorization of research activities, especially in fields like Humanities, that are constantly struggling for legitimacy.

---

[1] Notice the use of the word 'scientists' instead or 'researchers' or other related terms along the paper. In several academic and non-academic fora, we checked that the definition of both terms overlaps. Some experts consider that researchers can be scientists (scientific researchers) or not (researchers that do not follow the scientific method). Some others consider that scientists can be researchers (professional scientists) or not (anyone who follows a systematic way to increase understanding). Dictionaries are not much helpful, either. For instance, according to the Cambridge Dictionary, a researcher is someone engaged into the 'study of a subject in order to discover new information'; a scientist is 'someone who studies science or works in science', being science 'knowledge gained by observation and experiment'. We do not appreciate strong differences between the two terms, but we prefer 'scientist' because of its connotations ('passionate with what he/she does', 'motivated for the sake of knowledge'), which are slightly more positive than those of 'researcher'. Despite claims that Arts and Humanities are not science, we include scholars in these disciplines among scientists (we also considered the term 'scholar', but its definition according to the Cambridge Dictionary is 'a person of great knowledge and learning', which does not provide clearer cut points either).

The objectives of this paper are: (1) to place scientific authorship of literary fiction among existing taxonomies of knowledge transfer mechanisms, (2) to show the prevalence of scientific authorship of literary fiction, and (3) to analyze the relation between scientific authorship of literary fiction and knowledge transfer[2].

So far, we ponder knowledge transfer as to whether scientific fields of scientists are aligned with the contents of their fiction books. This refers to *direct knowledge transfer*. Our first research question is, 'do scientists use their literary works as mechanisms of direct knowledge transfer?' But this is not the only possible type of knowledge transfer. A scientist that publishes a fiction book enters into new cultural circles of editors, other writers, readers, etc. Consequently, scientists may spread their specialized knowledge in talks, presentations and other meetings. Regardless of the book contents and of whether it transfers scientific knowledge, its author would have more chances to let the public know about his/her research. We will call this *indirect knowledge transfer* and respond a second research question: do literary works give scientists the opportunity to transfer knowledge indirectly? In addition, we also consider that feedbacks may take place during the process of selling a book, including interactions from knowledge transfer to knowledge generation. The scientist, via contacts with the same cultural circles aforementioned, may benefit from the publishing experience through enriched life experiences, and thus get ideas to incorporate in his/her later research, which we call *reverse knowledge transfer*. This leads us to a third research question: do scientists experience flows from knowledge transfer to knowledge generation during the promotion of their literary work?

The paper continues with a conceptual section where we discuss the properties of literary fiction vis-à-vis other mechanisms of knowledge transfer, and we develop an initial framework for the analysis of the relation between scientific authorship of literary fiction and knowledge transfer. Then we analyze qualitative and quantitative evidences from two Spanish regions under this analytical framework. We discuss our finding and suggest propositions. Finally, we present our main conclusions.

---

[2] We consider literary fiction as the narrative forms of literature, i.e. novels, short stories, plays; thus leaving aside poetry or literary essay.

## 2 Literature review and conceptual approach

### 2.1 Scientific authorship of literary fiction and the distinction of formal versus informal knowledge transfer

Knowledge transfer is defined as the dissemination of professional knowledge from one person to another (OECD, 2000: 76). Based on such definition, we could assume, at least for this sub-section only, that all scientific authorship of literary fiction is a vessel of academic knowledge transfer. One way to discuss the characteristics of such mechanism is to place it within existing taxonomies of knowledge transfer channels. Because the property of literary fiction books is typically licensed to publishing companies, a taxonomy placing importance on intellectual property rights (IPRs) seems an appropriate departure point. Link et al. (2007) provide such IPR-centric taxonomy. They distinguish between formal and informal knowledge transfer mechanisms: formal refers to mechanisms that 'embody or directly result in a legal instrumentality such as, for example, patent, license or royalty agreement' (p. 642) and informal refers to mechanisms that foster 'the flow of technology knowledge but through informal communication processes, such as technical assistance, consulting and collaborative research' (p. 642). The later definition is a bit tautological, so Bradley et al. (2013) add that 'informal technology transfer is more abstract than formal technology transfer in that it involves the exchange of ideas and knowledge rather than the property of a specific invention' (p. 50). They also make a clear distinction between formal and informal knowledge transfer based on the disclosure or not of the scientific breakthrough to the university's technology transfer office (TTO). Bradley et al. (2013) also distinguish a third group of informal technology transfer channels, so-called 'academic-industry collaborations', which imply a sustained partnership over time, a 'working relationship' (p. 61) that may incorporate both formal and informal mechanisms, such as consulting, research contracts and joint labs. The conceptual boundaries between academic-industry collaborations and formal or informal knowledge transfer are not sharp yet, but this should have no effect on the inclusion of literary fiction in this framework, as we will now see.

Table 1 summarizes the above information and places scientific authorship of literary fiction among the different knowledge transfer channels. Writing literary fiction could be defined as formal knowledge transfer because it generates copyrights, a form of IPR. However, it does not pose the question whether to disclose it to the TTO or not, which qualifies it as informal knowledge transfer. Hence, scientific authorship of literary fiction does not fit perfectly into either formal or informal knowledge transfer. It could be then considered to form part of the group of 'academic-industry collaborations', as it combines elements of both. Nevertheless, we do not appreciate necessary signs of enduring partnerships between authors and editors that would support such classification.

**Table 1** Knowledge transfer types according to the formalization of their basic contents, including non-formal knowledge transfer

| Knowledge transfer type | Basic contents | Related mechanisms | Period of time |
|---|---|---|---|
| Formal knowledge transfer | Allocation of property rights, disclosure to TTO | Patents, licensing, spinoffs | Working time |
| Informal knowledge transfer | Communication processes, no disclosure to TTO, exchange of ideas | Talks and meetings, joint publications, technical assistance, free dissemination | Working time |
| Academic-industry collaborations | Sustained working relationship | Consulting, research contracts, joint labs | Working time |
| Non-formal knowledge transfer | Allocation of property rights, no disclosure to TTO | Non-academic books, copyrights, licensing | Spare time |

Source: (first three rows) own elaboration from Link et al. (2007) and Bradley et al. (2013)

We believe that the way to introduce scientific authorship of literary fiction is to consider it as a separate group. We propose the label of 'non-formal knowledge transfer' to refer to it. Non-formal knowledge transfer would be defined as the dissemination of professional knowledge embedded in planned activities that are not explicitly designated as a transfer mechanism, but which contain an important element of transfer[3]. The key to differentiate non-formal knowledge transfer from other knowledge transfer categories is to consider it as an activity for the spare time, whereas the rest are conducted during office hours. As a non-formal mechanism, scientific authorship of literary fiction involves management of IPRs (authors' license copyrights to publishing houses) but not the decision to disclose a result to the TTO as it lies beyond the control of the scientist's organization (it actually lies in the individual's private sphere).

## 2.2 Modeling the relationship between scientific authorship of literary fiction and knowledge transfer

Once having conceptualised scientific authorship of literary fiction as a non-formal knowledge transfer mechanism, we can dismiss now the assumption that all literary fiction involves scientific knowledge transfer.

---

[3] In the field of Educational Sciences, the distinction between formal, informal and non-formal learning or education is common (Colardyn and Bjornavold 2004). It has been our source of inspiration to propose the term and definition of 'non-formal knowledge transfer'.

Hence, we can speculate on its links with actual knowledge transfer. Fig. 1 tries to visualise the relations we aim to explore in this first approach. The following sub-sections tackle each one of these relations.

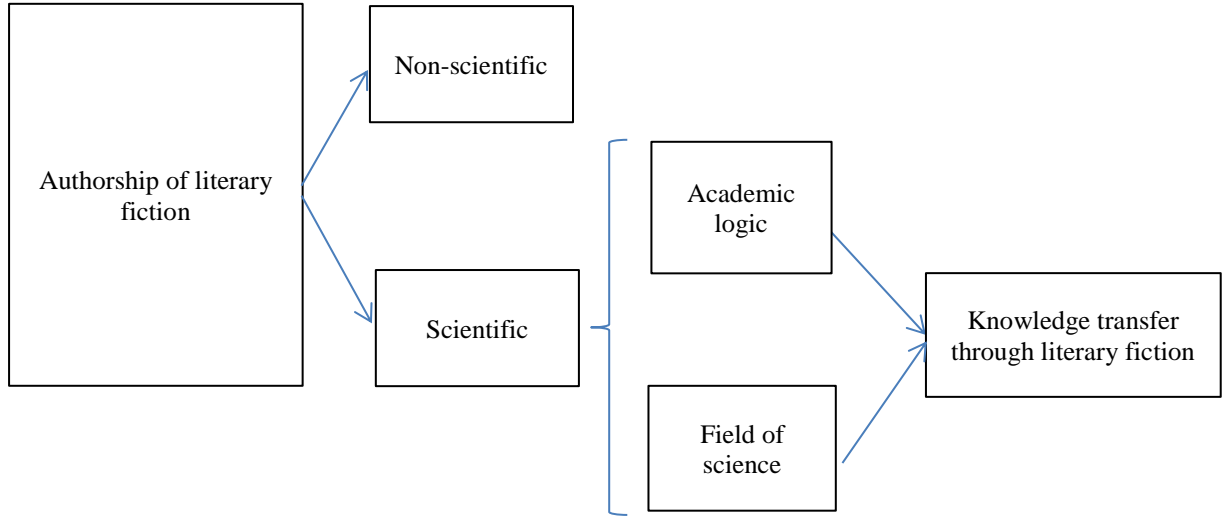

**Fig. 1** Conceptual framework of knowledge transfer via literary fiction

*2.2.1 Modeling scientific authorship of literary fiction*

Scientific production and university-society knowledge transfer depend on individual characteristics of researchers and other factors, e.g. age, gender, organization, scientific field, region, country, etc. (Carayol and Matt 2006; Baccini et al. 2014; Perkmann et al. 2013). Similarly, production of literary fiction depends on individual characteristics of the author and characteristics of the books and publishing companies, e.g. age, gender, genre, price, company type, etc. (Crozier 1999; Shehu et al. 2014; Thelwall 2017). At the interface of scientific activities and literary fiction production, we expect scientific authorship of literary fiction to be dependent on similar influences. More concretely, we may expect that these forces drive differences between scientific and non-scientific authors. In other words, scientific authors may represent a selected sample from the population of all authors, i.e. they may publish in distinct companies, dissimilar kind of books or simply have other personal characteristics. Hence, to analyze the contribution of scientists to knowledge transfer via literary fiction, we have to take into account scientific authorship of literary fiction as a previous step to the analysis of knowledge transfer via literary fiction. Our proposal is to identify the probability that an edited book of literary fiction is authored by a scientist:

$$SALF_{ijkl} = f(X_{ijkl}) \qquad (1)$$

Here, SALF is the probability that book i by author j, edition k, published by company l, has a scientific author. X is a vector of control characteristics, like book genre, geographic origin of the writer, company size, etc.

*2.2.2 Modeling direct knowledge transfer via literary fiction*

As mentioned in the introduction, we can differentiate three types of knowledge transfer: direct, indirect and reverse. This sub-section focuses on direct knowledge transfer (dissemination of scientific knowledge through the contents of the book) and the relationships with some of its driving forces. We propose the following model:

$$KT_{ij} = f(A_j, F_j) \qquad (2)$$

Where KT is the degree of knowledge transfer contained in a literary fiction work i by author j, A differentiates academic from non-academic scientists and F is a vector of the fields on which the scientists conduct research. We will deal with both aspects in more detail in the following sub-sections. We will recall indirect and reverse knowledge transfer, and further conceptualize the difference with direct knowledge transfer, in section 2.2.3.

*2.2.2.A Direct knowledge transfer via literary fiction according to the type of scientist (academic or non-academic)*

So far, we have dealt with academic scientists because of the relevance of their presence in the creative class of cities and the potential of literary fiction to transfer scientific knowledge. However, when we focus on this last aspect (knowledge transfer), we may think that it is not exclusive of academic scientists. Non-academic scientists can also transfer knowledge through their fiction books. We define academic scientists as research staff employed by universities and public research organizations (also known as government labs, e.g. the French CNRS, the Italian CNR, the Spanish CSIC, the German Max Planck Society, etc.). Non-academic scientists are defined as those employed by other organizations (e.g. industrial or hospital scientists), or who conduct research outside organizations (e.g. laypeople, citizen or independent scientists). Academic scientists have to compete with non-academic scientists, maybe in inferior conditions, in a similar vein to that of academic entrepreneurs who compete with non-academic entrepreneurs (Ayoub et al. 2016). We know that engagement in knowledge transfer activities may not give as much prestige as a research focus, and raise critical concerns for negative unintended consequences (Azagra-Caro et al. 2008; Perkmann et al. 2011). Similarly, academic scientists may be in inferior conditions than non-academic ones because under an academic logic, scientific authorship of literary fiction is not valued in the evaluation of merits, and peers may interpret knowledge transfer through literary fiction as frivolous. Writing fiction can even decrease academic scientists' reputation among scientific peers given that for these, the most valuable merit is publication in top journals (they may consider writing literary fiction as a waste of time, as suggested by some of our interviewees). On the contrary, under a non-academic logic, a literary fiction book may be an accepted form for scientists to disseminate expert knowledge and third parties may regard it as a complementary asset, given that third parties are not necessarily academics but societal actors who attach importance to literary fiction (most part of society does not read scientific articles). This would be apparent in the need for academic scientists to confine authorship of literary fiction to their spare time, whereas for non-academic scientists it may be part of their profession.

*2.2.2.B Direct knowledge transfer via literary fiction according to scientific fields: a knowledge transfer mechanism for the Social Sciences and Humanities?*

Knowledge transfer activities have different levels of intensity and use different mechanisms across scientific fields. Landry et al. (2006) found that Engineering followed by Earth Sciences were the areas with higher evidences of knowledge transfer, as opposed to Computer Science and Physics, Mathematics and Space Sciences. They also highlighted how commercialization does not always take place in these processes. Results are consistent with those reported by D'Este and Patel (2007). Such typical studies do not consider Social Sciences and Humanities in their analysis, implying that these fields have little capability of transferring knowledge to society. Another example of such is the study by Tijssen (2006), which does not include any of these fields among what he calls 'research areas of significant industrial interest'.

However, recently, some studies have suggested that scientists in Social Sciences and Humanities simply employ other knowledge transfer mechanisms. Examples of such studies are those by Cassity and Ang (2006) or by Olmos-Peñuela et al. (2014). Strikingly, this latter study reported levels of relational knowledge transfer activities in Social Sciences and Humanities similar to those reported in Engineering and the Natural and Physical Sciences, through mechanisms like consultancy or contract research.

Molas-Gallart and Tang (2011) found that researchers from the fields of the Social Sciences use indirect mechanisms of communication –that is, through text– as a means to interact with non-academic stakeholders. And indeed, it is a means with the potential to lead to further direct interactions. This is intimately related with the purpose and forms of communication followed in Social Sciences and Humanities where there is no strict line between what is directed to the scholarly community and what is directed to the public. Here it is important to mention the audience framework developed by Nederhof (2006), where he considers that researchers in these fields may publish their research results in different forms depending on their target audience. He distinguishes three potential audiences: other scientists, non-scholar experts and the general audience. In this regard, the distinction between what is research output, what is intended to communicate and disseminate science among lay citizens and what is clearly targeted towards transfer knowledge becomes blurred.

In this sense, we might expect a higher degree of knowledge transfer via literary fiction in fields such as Social Sciences and Humanities than suggested by other mechanisms, and higher than in other scientific fields.

### 2.2.3 Indirect and reverse knowledge transfer

The discussion up to now has been mainly about *direct knowledge transfer* (DKT), which evokes the embodiment of scientific knowledge on a mechanism (the book in this case) that mediates between the scientist and the recipient. We can distinguish it from other forms of knowledge transfer, according to the codification and direction of the knowledge flow. DKT would be codified in the form of the book and directed from the scientist to the reader.

Contributions of universities to their surroundings can be also more indirect (Smith and Bagchi-Sen 2012). In the process of diffusion of a book, a scientist may have other opportunities to transfer knowledge, namely by word-of-mouth with editors, other writers, attendants to presentations and meetings, and the readers, i.e. the cultural world. We call this *indirect knowledge transfer* (IKT), which is characterized by the presence of tacit knowledge and still directed from the scientist to the cultural world.

**Table 2** Knowledge transfer types via literary fiction according to the codification and direction of related knowledge

| Knowledge transfer type | Basic contents | Knowledge type | Direction of the knowledge flow | Empirical analysis in: |
|---|---|---|---|---|
| Direct knowledge transfer | The book includes the scientist's research topics | Codified | Scientist-Public | Section 4.2 (interviews and regression) |
| Indirect knowledge transfer | Scientific authors talk about their research with cultural agents | Tacit | Scientist-Public | Section 4.3 (interviews) |
| Reverse knowledge transfer | Cultural agents give scientists ideas for future research | Tacit | Public-Scientist | Section 4.3 (interviews) |

We know also that knowledge transfer can run from industry to science (Fier and Pika 2014). Similarly, in contact with this cultural world, the scientist may obtain ideas for future knowledge generation, and thus generate a feedback from literary fiction to scientific production. This can be considered *reverse knowledge transfer* (RKT), based on tacit knowledge and directed from the cultural world to the scientist. Table 2 summarizes these ideas.

At the left-hand side of Fig. 1, we had scientific authorship of literary fiction, which can be analysed for each type of knowledge transfer: direct, indirect and reverse. In the following section, we are going to develop a qualitative methodology to inform and refine our theory about DKT as a function of academic logic and field of science (middle-box of Fig. 1 and right hand-side of Equation 2), and a quantitative methodology to test some of the most simple relationships. Hence, through evidence from interviews to scientific writers and estimated regressions of Equations 1 and 2, we will answer the first research question mentioned in the introduction, referred to DKT ('do scientists use their literary works as mechanisms of direct knowledge transfer?'). Then, we will use the interviews to inform and refine our theory about IKT and RKT and answer our second and third research questions ('do literary woks give scientists the opportunity to indirectly transfer knowledge?', 'do scientists experience reverse flows from knowledge transfer to generation in the diffusion of their literary fiction?'). The last column in Table 2 relates each type of knowledge transfer to the section where the analysis is done and through which method.

## 3 Methodology and data

Data to answer questions about scientific authorship of literary fiction and knowledge transfer were extracted from the Spanish national database of books published by the ISBN Agency[4]. The Spanish Ministry of Culture hosts this database along with a national publishers' database. We included only books published in 2015. We selected this year for its particular interest. In the global publishing market, market value had been decreasing from 2006 until 2015, but rose in 2015 and the expectation is of a sustained increase in the following years (Marketline 2016). This upward shift in the trend is true at global level and, as Fig. 2 shows, in the particular case of Spain and of two Spanish regions: the Valencian Community and Catalonia. These communities were selected because both are representative in terms of revenue trend of the rest of Spain and due to their geographical location in the periphery of Spain.

---

[4] http://www.mecd.gob.es/cultura-mecd/areas-cultura/libro/bases-de-datos-del-isbn.html

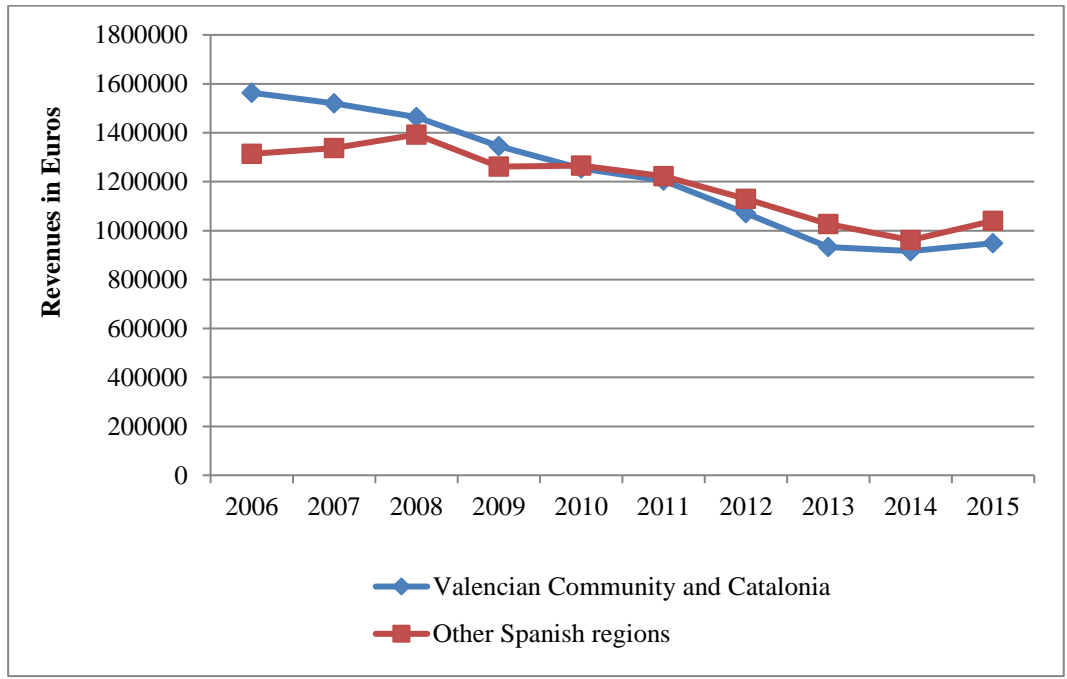

**Fig. 2** Evolution of total revenues of Spanish publishing companies. Source: own elaboration based on the Spanish version of Bureau Van Dick's Orbis, i.e. the Iberian Balance Analysis System (SABI).

We first extracted a list of publishers based in these communities and then searched in the ISBN database for their output in 2015. In the case of the Valencian Community (all publishers), we expected to find small publishing companies. In the Catalan case (60% of publishers in terms of revenues), we found that this region comprises the main cluster of publishing companies in the country. Hence the size and company type are of a complete different scale in each region. The same can be said about the presence of higher education and research institutes. Barcelona comprises 12 universities, 8 of which are public and of a large size such as University of Barcelona or Autonomous University of Barcelona. In the case of the Valencian Community, we find 9 universities, 5 of which are public and 4 private. These dissimilarities are partly due to the differences on population size. While Catalonia is the second Spanish community with the largest population (7.5 million people approximately), the Valencian Community has around 5 million people. This gives a variety of cases, ranging from local, small companies, which publish 'amateur' writers along with large, international companies publishing bestsellers. Our companies published 3,765 books in 2015. By manually checking online information about the books, we configured a valid sample of 541 fiction works (a 14% of the total). Fig. 3 helps visualizing the structure of the database.

To build our measure of SALF, we identified whether the authors of the fiction works were scientists. A well-reported issue when working with monographs is the lack of address information of the authors (Gorraiz et al.

2013). This makes it problematic when trying to identify the institutions behind such works or the background of the author. Here, the identification of scientists was done manually by checking in the Internet and searching for information on the author's affiliation and the presence of evident signs of having done research, e.g. having completed a PhD, having won research prizes, etc.

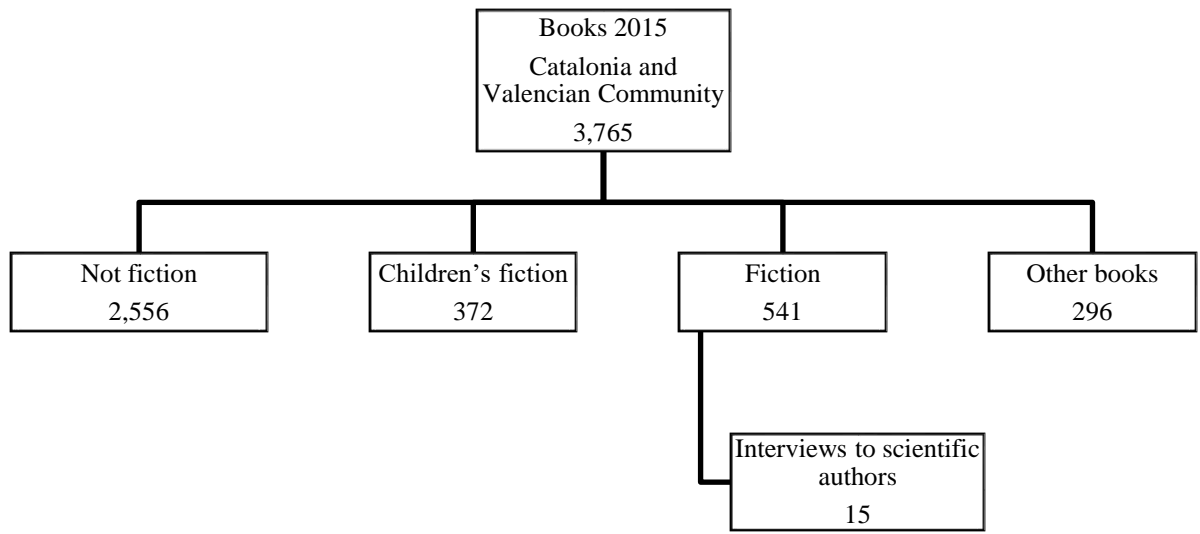

**Fig. 3** Identifying literary fiction books in Catalonia and the Valencian Community (2015)

Fig. 4 shows that 11% of authors were scientists. This is also the mean of our measure of SALF. We distinguished between academic (working at a university or at a public research organisation) and non-academic scientists. We identified the latter by searching for evidences of a research background, such as working at an industrial R&D lab, having done a PhD, having won a research prize or having authored a research paper. Fig. 4 shows that our sample of scientists is distributed in favour of academic scientists, although not overwhelmingly.

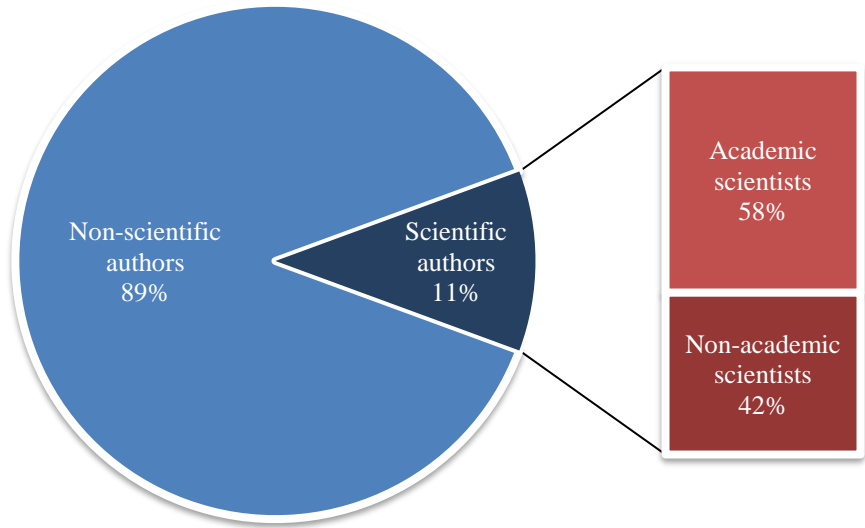

**Fig. 4** Scientific authors in Catalonia and the Valencian Community in 2015

Table 3 offers details about the criteria used for individuals to qualify within each category. Our sample includes historical figures like Ramón Llull (1232-1315), a philosopher; Johann Wolfgang von Goethe (1749-1832), author of 'Faust' and a natural scientist who published treaties in Botanics, Zoology and Optics; or Arthur Conan Doyle (1859-1930), the creator of Sherlock Holmes, who published papers in Medicine. It also includes international bestselling scientific writers like Deepak Chopra or Diana Gabaldón, and Spanish national counterparts like María Dueñas or Joaquín Camps.

To build DKT1, for scientific authors, we read their books' plot summaries and additional information where necessary. We considered that their fictional work transfers the contents of their research if the plot or theme is related to the scientist's research lines. The corresponding binary variable DKT1 takes value 1 if there is knowledge transfer and 0 otherwise. Our rating method to feed DKT is a superficial approach to the book contents compared to reading the whole book, but it increases feasibility. Four people gathered the data and rated some of the books independently, based on summaries and other information available. The instruction given was, 'Is the plot related to the author's research lines?'[5] A fifth person who did not gather the data acted as referee and rated all the books to provide a more homogenous rating. A sixth one acted as second referee in case of (very few) discrepancies. We detected the presence of direct knowledge transfer in 32% of the books with scientific authors. Hence, we observe that some authors made accessible their research interest to the public through literary fiction books. This is in line with the study of Nederhof (2006), who defended that some scientists publish their research results in different forms to arrive to a broad audience.

---

[5] We have also generated a variable on *contextual* knowledge transfer, i.e. whether in fiction works the action occurs in a scientific setting, prominent characters are scientists, etc., but we left it out of the analyses for simplification.

**Table 3** Distribution of scientific author types

| Scientific author type | Number | % |
|---|---|---|
| **Academic scientist** | **36** | **58%** |
|     Public research organization scientist | 3 | 5% |
|     University scientist | 33 | 53% |
| **Non-academic scientist** | **26** | **42%** |
|     Industrial scientist with a PhD or who has occasionally worked at university or published scientific reports | 3 | 5% |
|     Hospital scientist with a PhD | 1 | 2% |
|     Ex academic | 2 | 3% |
|     Has a PhD and may have occasionally worked at university and/or published in scientific journals | 7 | 11% |
|     Has presented in academic conferences | 1 | 2% |
|     Has published in scientific journals, proceedings, books or reports | 11 | 17% |
|     Has won research prizes | 1 | 2% |
| **Total** | **61** | **100%** |

Concrete examples of what we considered presence of knowledge transfer are Tim Bruno, a marine biologist with a novel in which the main character and his friends sail the sea of a fantastic world; Christian Jacq, an Egyptologist who has been writing fiction and non-fiction books about Egypt for long; or, at a more local level, Isabel Canet Ferrer, a historian with expertise on the Monastery of Valldigna, which also appears in her novel.

Apart from SALF and DKT1, we built other variables for the regressions. Table 4 shows the list, and their descriptive statistics. We classified all scientific authors into five scientific fields based on the OECD (2007) subject classification: Natural Sciences, Medical and Health Sciences, Social Sciences, Linguistics and Literature and Other Humanities[6]. If one author had more than one field, we duplicated the number of observations and tabulated if knowledge transfer occurred in any of the fields. Most scientists in our sample belong to Linguistics and Literature, and Social Sciences.

---

[6] We made this distinction within Humanities because of the large number of authors that belonged to Linguistics and Literature, and for the unusual patterns of knowledge transfer in Linguistics and Literature coming out of the interviews (see section 4.2.2).

**Table 4** List of variables and descriptive statistics

| Equation | Level | Name | Description | Mean | S.D. | Minimum | Maximum |
|---|---|---|---|---|---|---|---|
| 1 | Characteristics of the book (dependent variable) | Scientific authorship of literary fiction (SALF) | The book's author is a scientist | 0.11 | 0.32 | 0 | 1 |
| | Characteristics of the book (independent variables) | Mode | 1=Novel, 0=Short Stories, Drama, Aphorisms… | 0.92 | 0.24 | 0 | 1 |
| | | Non-genre book | | 0.23 | 0.42 | 0 | 1 |
| | | Genre 1 | Crime-Mystery-Noir-Thriller-Horror | 0.19 | 0.39 | 0 | 1 |
| | | Genre 2 | Fantasy-Science fiction | 0.06 | 0.24 | 0 | 1 |
| | | Genre 3 | Historical-Adventure | 0.16 | 0.36 | 0 | 1 |
| | | Genre 4 | Romantic-Erotic | 0.16 | 0.37 | 0 | 1 |
| | | Genre 5 | Young adult fiction (benchmark) | 0.20 | 0.40 | 0 | 1 |
| | Characteristics of the author | Gender | 1=Female, 0=Male (benchmark) | 0.43 | 0.50 | 0 | 1 |
| | | Author´s year of birth | | 1947 | 75.75 | 1232 | 1996 |
| | | Region 1 | Author from Valencian Community | 0.19 | 0.39 | 0 | 1 |
| | | Region 2 | Author from Catalonia | 0.09 | 0.28 | 0 | 1 |
| | | Region 3 | Author from other regions of Spain | 0.26 | 0.44 | 0 | 1 |
| | | Region 4 | Foreign authors (benchmark) | 0.46 | 0.50 | 0 | 1 |
| | Characteristics of the edition | Format | 1=E-book, 0=Printed (benchmark) | 0.21 | 0.40 | 0 | 1 |
| | | Price | In euros | 11.63 | 5.94 | 0.82 | 40.38 |
| | Characteristics of the publishing company | Size 1 | Largest company | 0.53 | 0.50 | 0 | 1 |
| | | Size 2 | Second largest company | 0.14 | 0.35 | 0 | 1 |
| 2 | Characteristics of the book (dependent variable) | Direct knowledge transfer (DKT1) | The book includes the scientist's research topics | 0.32 | 0.47 | 0 | 1 |
| | Characteristics of the scientific author | Scientific author type | 1= Academic  0= Non academic | 0.58 | 0.50 | 0 | 1 |
| | | Scientific field 1 | Natural Sciences | 0.07 | 0.25 | 0 | 1 |
| | | Scientific field 2 | Medical and Health[7] | 0.12 | 0.33 | 0 | 1 |
| | | Scientific field 3 | Social Sciences | 0.29 | 0.46 | 0 | 1 |
| | | Scientific field 4 | Linguistics and Literature | 0.38 | 0.49 | 0 | 1 |
| | | Scientific field 5 | Other Humanities (benchmark) | 0.15 | 0.36 | 0 | 1 |

---

[7] It includes one observation from Engineering and Technology.

We will use our quantitative data about DKT after informing our theory with qualitative data out of interviews with scientific authors. Between September and October 2016, we conducted interviews to 15 scientific writers from our sample. In an introductory mail or call, we verified in the first place whether we had identified interviewees correctly as 'scientists'. All considered themselves to be conducting or having conducted scientific research, including those in Humanities. The interviews (20-30 minute long) included questions about the research line of the respondents, the publishing history of their book, the relation between their research line and their book, opportunities raised due to having published fiction literature to talk about their research lines with new audiences, and how this stimulated their own research. The fiche of individuals can be found in Table 5.

**Table 5** Interview data

| Interviewee code | Scientific field | Scientific sub-field |
| --- | --- | --- |
| Academic scientist 1 | Social Sciences | Economics |
| Academic scientist 2 | Humanities | Linguistics and Literature |
| Academic scientist 3 | Social Sciences | Economics |
| Academic scientist 4 | Humanities | Linguistics and Literature, Philosophy |
| Academic scientist 5 | Humanities | Linguistics and Literature |
| Academic scientist 6 | Social Sciences | Media and Communications |
| Academic scientist 7 | Humanities | Linguistics and Literature |
| Academic scientist 8 | Social Sciences | Business |
| Academic scientist 9 | Humanities | History |
| Academic scientist 10 | Social Sciences | Educational Science[8] |
| Academic scientist 11 | Humanities | History |
| Non-academic scientist 1 | Social Sciences and Humanities | Linguistics and Literature, Educational Science |
| Non-academic scientist 2 | Humanities | Linguistics and Literature |
| Non-academic scientist 3 | Humanities | Linguistics and Literature |
| Non-academic scientist 4 | Medical and Health Sciences | Clinical medicine |

Names of scientific fields are taken from OECD (2007).

Interviews to scientific writers included open questions about self-evaluations of direct knowledge transfer (DKT2), indirect knowledge transfer (IKT) and reverse knowledge transfer (RKT). Two individuals coded the answers in respective High/Low scales, indicating presence/absence of such transfer and a third one refereed in case of discrepancy (just one). The second column of Fig. 5 provides the measure obtained for DKT2 (50%). Self-reported direct knowledge transfer tends to be higher than externally evaluated direct knowledge transfer (DKT1=32%). The third and fourth columns of Fig. 5 report 43% and 33% values for indirect and reverse knowledge transfer, respectively.

---

[8] During the interview, we realized Academic scientist 10 had joined academia after having written her novel, so we removed her from the analysis of direct and indirect knowledge transfer. However, she is especially useful for the analysis of reverse knowledge transfer, since the risk of endogeneity is lower than in cases of having written the book during the professional practice.

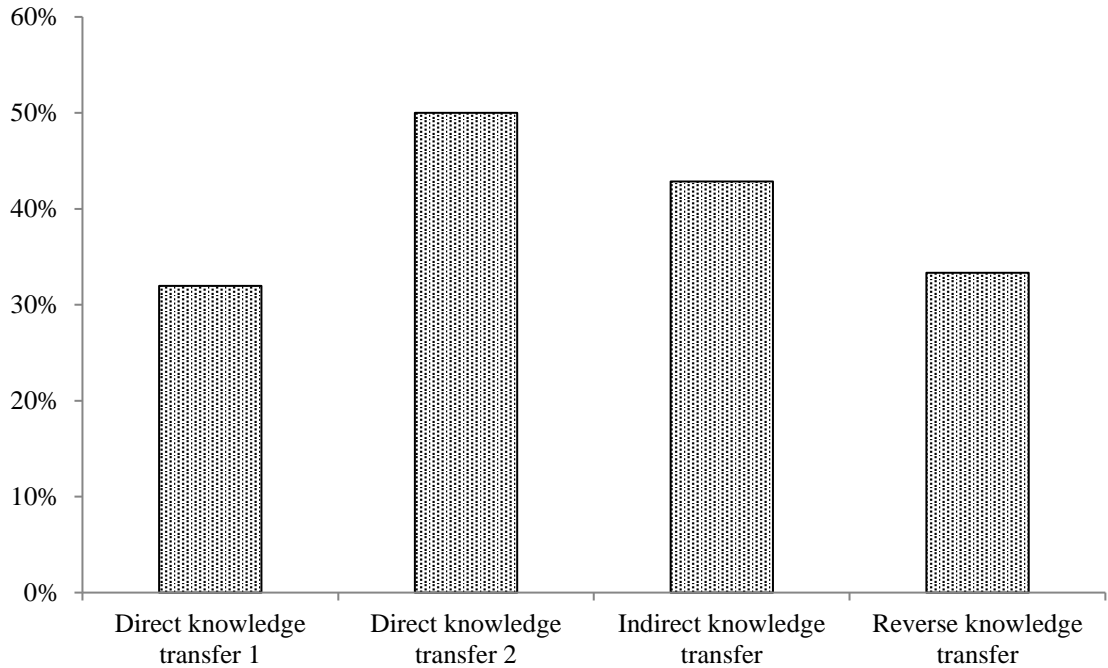

**Fig. 5** Knowledge transfer via literary fiction in Catalonia and the Valencian Community (2015)

In the following section, we present our results grouped by insight, broken down by sub-sections that mirror those from section 2.1.

## 4 Results

### 4.1 Scientific authorship of literary fiction

In section 2.2.1, we anticipated that scientific authorship of literary fiction would be dependent on individual characteristics of the authors and of the type of book and publishing company. We estimate Equation 1 through a probit model, which is adequate because the dependent variable is binary (the author is either a scientist or not). Despite the use of econometrics, we are not testing concrete hypotheses but rather using the method to provide exploratory evidence of significant statistical relationships that could serve as basis for future research. This is because the focus of the paper is on knowledge transfer via literary fiction and on the relationship between this transfer and scientific authorship of literary fiction. Hence, the interest in scientific authorship of literary fiction is not per se but conditional on its relationship with knowledge transfer.

Table 6 contains the results and the chi-square confirms that the model is significant. With regards to the characteristics of the book, curiously, being a novel (as opposed to other forms of fiction) is negatively significant, which means that scientists tend to write short stories –an explanation for this phenomenon could be that writing short stories is in line with the concision of the scientific writing style. Scientific authors tend to publish more in the historical-adventure genre.

**Table 6** Probit estimation of scientific authorship of literary fiction

| Variable | Coefficient |
| --- | --- |
| Novel | -0.49** |
|  | (0.22) |
| Non-genre book | -0.01 |
|  | (0.20) |
| Crime-Mystery-Noir-Thriller-Horror | 0.20 |
|  | (0.18) |
| Fantasy-Science fiction | 0.16 |
|  | (0.30) |
| Historical-Adventure | 0.38** |
|  | (0.19) |
| Romantic-Erotic | -0.29 |
|  | (0.28) |
| Young adult fiction (benchmark) |  |
| Female author | 0.04 |
|  | (0.15) |
| Author´s year of birth | -0.00 |
|  | (0.00) |
| Author from Valencian Community | 0.32* |
|  | (0.19) |
| Author from Catalonia | -0.44* |
|  | (0.25) |
| Author from other regions of Spain | 0.03 |
|  | (0.13) |
| Foreign author (benchmark) |  |
| E-book | 0.48*** |
|  | (0.17) |
| Price | 0.03*** |
|  | (0.01) |
| Largest company | -0.40** |
|  | (0.18) |
| Second largest company | -0.29 |
|  | (0.20) |
| Constant | 10.03*** |
|  | (2.27) |
| Observations | 541 |
| $\chi^2$ | 24.11 |
| p | 0.00 |

\* $p<0.1$; ** $p<0.05$; *** $p<0.01$. Standard errors in brackets. Weighted by inverse number of duplications due to multiple authoship, field of science or format.

About characteristics of the author, age and gender do not have an influence, but the probability of finding scientific authors increases when the author is from the Valencian Community and decreases when the author is from Catalonia. This may be due to differences in the characteristics of publishers in each region as discussed above.

Characteristics of the edition are the most significant. Scientific authors tend to publish more e-books (rather than printed versions) than non-scientific authors, and the price of scientific authors' books is higher than the price of books published by non-scientific authors. About the publisher, larger companies publish fewer books written by scientific authors than the rest.[9]

We will use the former estimation as Step 1 of a Heckman probit model of the presence of direct knowledge transfer according to type of scientist and scientific field. Before proceeding to do so, we will use our qualitative evidence to increase understanding of the channels involved in the relationships and refine our conceptual theory from section 2.2.2.

## 4.2 Direct knowledge transfer

### *4.2.1 Direct knowledge transfer and academic logic: qualitative results*

In section 2.2.2.A, we speculated whether academic scientists were less prone to transfer knowledge directly than non-academic scientists. Our qualitative findings tend to support this view. Table 7 shows that, according to our coding of respondents' answers, the percentage of non-academic scientists providing oral evidence of direct knowledge transfer in their books is higher than that of academic scientists.

**Table 7** Distribution of interviewees according to scientist type and direct knowledge transfer (n=14)

|  | Low direct knowledge transfer (DKT2=0) | High direct knowledge transfer (DKT2=1) | Total |
|---|---|---|---|
| Academic scientist | 60% | 40% | 100% |
| Non-academic scientist | 25% | 75% | 100% |
| Total | 50% | 50% | 100% |

---

[9] As deducible from historical examples in section 3, we have included deceased writers in the sample. This was important conceptually to show the prevalence of the phenomenon of scientific writers. However, there are not many deceased writers (9%) and their inclusion does not condition the results. When we include a dummy for deceased authors or remove them from the sample, the sign and significance of all coefficients hold.

Some extracts from interviews about academic scientists' motivations for writing literary fiction illustrate why so few knowledge transfer in their case. The typical academic scientist declares that his/her fiction work is not about his/her research topics, so there is no knowledge transfer:

*It's like a form of evasion –you get to like it, and what starts being a hobby […] is finally like a second job.*
(Academic Scientist 3)

The terms 'hobby' and 'evasion' or synonyms are recurrent in interviews. However, the academic profession is supposed to be creative and endow much freedom. Why such a need for evasion? One reason may be the perception that *scientific* creativity is at odds with *artistic* creativity:

*The use of language with aesthetic and artistic goals is something you cannot afford […] when you're writing academic works and papers.*
(Academic Scientist 5)

This resonates with the idea that artistic and scientific personalities are different: artistic personality is, for instance, more reluctant to norms (Feist 1998). To the very end, literary fiction may be more creative and motivating than research, so lack of organizational and institutional support to literary fiction may translate into employee burnout (Madjar 2008):

*Professional accomplishment of academic routines […] stopped being satisfactory.*
(Academic Scientist 6)

Sometimes, alleviating the tension from this misalignment becomes a practical concern, which makes sense given that growth need strength motivates creativity (Shalley et al. 2009):

*[The academic and literary world] are very apart from each other […]. I'd love there were bridges, channels from one to another […]. My dream would be to try to link research with […] things that I can use in my stories, and not to be constantly schizophrenic […].*
(Academic Scientist 11)

The same individual exemplifies to what extent alleviating the disconnection can have a positive effect on personal development:

*People from my academic world have been invited to presentations [...], so it has forced me to show a facet of mine that was not known at public level: it's been like 'getting out of the closet', the fact of saying, 'Well, I also do this'. And this has been useful, and good to me, and has made me lose some inhibitions.*

By opposition to the above quotes, non-academic scientists gives concrete examples of knowledge transfer of their research lines in their fiction works, e.g.:

*[For my work] I did a lot of research on bullying [...] and thought about writing an essay on this [...]. Why not a novel for teenagers to work out the problem of school bullying? [...]. The police from [a town] saw this book, and proposed it at every institute and I went to all of them to talk about the book.*
(Non-Academic Scientist 1)

As anticipated in section 2.2.2.A, we attribute this alignment to the usefulness of literary fiction for competence signalling in non-academic scientists. Other examples can be found in the Medical and Health Sciences:

*I like scientific knowledge within the work and in all my novels there is some aspect of scientific dissemination. Even in [the novel from the sample], on Forensic Medicine and Psychology. There is a character that has an obsessive-compulsive disorder, another one who is autistic…*
(Non-Academic Scientist 4)

Hence, our qualitative findings inform the implicit proposition from our conceptual theory that academic logic reduces direct knowledge transfer via literary fiction, and suggest explanations rooted in organizational and management psychology. We now move to the relationship between this transfer and scientific fields.

*4.2.2 Direct knowledge transfer and scientific fields: qualitative results*

In section 2.2.2.B, we speculated whether literary fiction gave Social Sciences and Humanities the opportunity to stand out in knowledge transfer compared to other scientific fields, contrary to what happens when comparison is based on usual knowledge transfer channels such us patents, R&D contracts, spinoffs, etc.

We selected informants from those fields mainly, to increase particular understanding about knowledge transfer via literary fiction there. Three patterns emerged: (1) in Social Sciences, contrary to our expectations, there was no much knowledge transfer; (2) in Humanities, there was considerable knowledge transfer; (3) within Humanities, the case of Linguistics and Literature was different from that of Other Humanities (mainly History and Philosophy). More specifically, whereas in Other Humanities our external perception of direct knowledge transfer (DKT1) matched the authors' own perception (DKT2), this was not true for Linguistics and Literature.

Table 8 gives a detailed account of such relation. Before interviews, we considered that most scientific writers interviewed did not transfer knowledge in their books (DKT1=0). After interviews, we confirmed this choice in less than half of the cases (DKT2=0) but not for the other half (DKT2=1). A quick look at the data suggested that the lack of validation affected mostly to scientists working in the fields of Linguistics and Literature. Even out of the two scientific writers interviewed that we considered transferring knowledge in their books (DKT1=1), one did not consider transferring that much (DKT2=0). He also belonged to the field of Linguistics and Literature.

**Table 8** Distribution of interviewees according to external (DKT1) and internal (DKT2) perception of direct knowledge transfer

|  | DKT2 = 0 | DKT2 = 1 | Total |
| --- | --- | --- | --- |
| **DKT1 = 0** | **6** | **6** | **12** |
| Linguistics and Literature | 1 | 5 | 6 |
| Other scientific fields | 5 | 1 | 6 |
| **DKT1 = 1** | **1** | **1** | **2** |
| Linguistics and Literature | 1 |  | 1 |
| Other scientific fields |  | 1 | 1 |
| **Total** | **6** | **7** | **14** |

The explanation of the divergence between DKT1 and DKT2 in Linguistics and Literature may lie in the scarce distance between research objects and the fiction book as a transfer mechanism in this field. For instance, linguists may study structural or stylistic properties of language that are embedded in the book, but become hermetic to the external evaluator of knowledge transfer. Scholars in the field are disseminating knowledge about the use of language through their fiction books. We call this 'format-content interference' and it only affects Linguistics and Literature. Because of this, DKT1 could underestimate direct knowledge transfer via literary fiction in this field. Belonging to the field of Linguistics and Literature may increase direct knowledge transfer via literary fiction, but in ways that are not so tangibly disseminated to the reader as in Other Humanities.

Our qualitative findings have two implications: First, we need to reformulate the implicit proposition in section 2.2.2.B that Social Sciences and Humanities stand out in knowledge transfer via literary fiction; now, we expect only Humanities and not Social Sciences to stand out. Second, for an econometric validation of a positive relationship between Humanities and direct knowledge transfer measured via an externally evaluated indicator of direct knowledge transfer such as DKT1, we have to split Linguistics and Literature from Other Humanities, since DKT1 is not likely to capture direct knowledge transfer in Linguistics and Literature. Hence, we expect to validate through econometrics a positive relationship between DKT1 and Other Humanities only. This differentiation also ratifies the decision to split Linguistics and Literature from Other Humanities taken after the large number of individuals in Linguistics and Literature in the sample (see footnote 6).

*4.2.3 Direct knowledge transfer, academic logic and scientific fields: Quantitative results*

The analysis of interviews has opened up some of the channels behind the relationships between direct knowledge transfer and two of its potential driving forces: academic logic and scientific field, as proposed in section 2.2. We can now proceed with some quantitative analysis of the most basic relationships, i.e. direct effects of academic logic and scientific field on direct knowledge transfer. This loses the richness of qualitative data but offers a first methodological approach to explore systematically samples of larger size that those used for the qualitative insights.

We start with some simple statistics. Table 9 includes the average value of DKT1 according to whether scientific authors are academics or not. We can see that non-academic scientists tend to transfer knowledge more often than academic scientists do via literary fiction.

**Table 9** Average direct knowledge transfer by scientific author type and field (n=61)

| Variable | Dimension | Average direct knowledge transfer (DKT1) |
|---|---|---|
| Scientific author type | Academic | 0.24 |
| | Non-academic | 0.43 |
| Scientific field | Natural Sciences | 0.25 |
| | Medical and Health Sciences | 0.60 |
| | Social Sciences | 0.23 |
| | Linguistics and Literature | 0.11 |
| | Other Humanities | 0.83 |

Weighted by inverse number of duplications due to multiple authoship, field of science or format.

Table 9 also includes the average value of DKT1 by scientific field. Specifically, we find that scientific authors of literary fiction from Other Humanities and Medical and Health Sciences publish about topics related to their fields above the sample average (0.32 according to Table 4), whereas the rest of fields are under average. This includes Social Sciences and Linguistics and Literature, as suggested by our qualitative findings.

These results need to be tested in an econometric setting, to control for sample selection and other factors. This is what we do next. We run a Heckman probit model that uses estimates of equation 1 from Table 6 in the first step and equation 2 in the second step. Table 10 presents the results of this second step. The Heckman model is a priori appropriate because there might be issues of self-selection bias that could influence the findings about knowledge transfer (namely, if scientific authors are not standard authors). For instance, we could expect a larger number of scientific authors in the historical genre because of the codes of that genre (history scholars are better documented about the context of the action than other authors). This is why equation 1 acts as a selection equation for equation 2. A posteriori, the significant chi-square of the selection equation in Table 10 indicates that there is evidence of sample selection, so it confirms that the Heckman probit model is adequate.

**Table 10** Heckman probit estimation of presence of direct knowledge transfer in literary fiction works: Step 2

| Variable | Coefficient |
| --- | --- |
| Academic scientist | -0.66** |
|  | (0.30) |
| Natural Sciences | -1.66*** |
|  | (0.55) |
| Medical and Health Sciences | -0.90*** |
|  | (0.34) |
| Social Sciences | -1.34*** |
|  | (0.39) |
| Linguistics and Literature | -1.59*** |
|  | (0.34) |
| Other Humanities (benchmark) |  |
| Constant | -0.26 |
|  | (0.40) |
| Observations | 541 |
| Censored observations | 480 |
| $\chi^2$ | 48.20 |
| p | 0.00 |

* $p<0.1$; ** $p<0.05$; *** $p<0.01$. Step 1 is the estimation in Table 6. Standard errors in brackets. Weighted by inverse number of duplications due to multiple authoship, field of science or format.

According to the results, we endorse most findings from Table 9. In Table 10, on the one hand, we see that the presence of knowledge transfer is negatively related to the scientific author being academic. Hence, we observe that non-academic scientific authors (those who are not working at a university or at a public research organisation) transfer knowledge through their literary works more often than academic authors do. On the other hand, belonging to Other Humanities increases the probability of finding knowledge transfer, compared to other fields of science. Medical and Health Sciences, despite scoring above average in DKT1 (Table 9), do not stand out vis-à-vis Other Humanities, once sample selection and academic logic are accounted for.

**4.3 Indirect and reverse knowledge transfer**

In section 2.2.3, we claimed that the analysis of forms of knowledge transfer other than direct knowledge transfer was important to provide a comprehensive picture of literary fiction by scientific authors. Our qualitative evidence let us be more specific, and disentangle some relationships between indirect/reverse knowledge transfer and academic logic, analogous to the relationship between direct knowledge transfer and academic logic, but pointing in another direction.

Let us start with *indirect* knowledge transfer. Our coding of interviewees' responses on IKT (i.e. via contact with cultural circles), appears in Table 11. Compared to DKT in Table 7, almost the opposite picture arises: academic scientists engage more often on IKT than non-academic scientists.

**Table 11** Distribution of interviewees according to scientist type and indirect knowledge transfer (n=14)

|  | Low indirect knowledge transfer (IKT2=0) | High indirect knowledge transfer (IKT2=1) | Total |
|---|---|---|---|
| Academic scientist | 40% | 60% | 100% |
| Non-academic scientist | 100% | 0% | 100% |
| Total | 57% | 43% | 100% |

Apparently, it is commonplace for academic scientists to discuss their own research topics with, for example, editors:

*My research work is something that I've talked about many times in informal conversations with my editors […] because they were interested in the topic I had worked through; they got my essay book […]. Hence, they are always informed about that work and […] they value it very positively.*

(Academic Scientist 5)

Occasionally this indirect transfer yields academic benefits for the author (and further engagement into transfer activities):

*After having published with them, [the editors] called me to do the foreword of a book by [another academic scientist], which is a book about sustainable economy [...]. It amused them, didn't it? 'We have a writer who knows about Economics and who is professor of Economics on top'.*

(Academic Scientist 3)

Hence, editors' awareness of expert knowledge by academics is likely to enhance the positive influence of academic logic on indirect knowledge transfer. This is logical given the importance of editors for selecting contents (Banou 2013). In addition, informal contacts may reach beyond editors:

*Conversation [with the public] may derive to, 'What is your research about?' So… 'university-industry relationships' [...] and this is a topic about which people have an opinion [...]: 'It's good that universities work for companies because otherwise universities study useless things' [or, on the contrary] 'It's bad that universities work for companies because companies have their own interests' [...]. This way you start a dialogue with people [...] and I can go a bit deeper into the results of my research.*

(Academic Scientist 1).

Therefore, selection of topics with some degree of public controversy seems to be a key driver of indirect knowledge transfer. On the opposite side, non-academic scientists did not recall any similar memory about this type of transfer.

Let us now continue with *reverse* knowledge transfer. In Table 12, we can see that academics also practice it more often than non-academics. Reverse knowledge transfer is just not as frequent as indirect knowledge transfer for academics is, whereas for non-academics both are equally rare.

**Table 12** Distribution of interviewees according to scientist type and reverse knowledge transfer (n=15)

|  | Low reverse knowledge transfer (RKT2=0) | High reverse knowledge transfer (RKT2=1) | Total |
|---|---|---|---|
| Academic scientist | 55% | 45% | 100% |
| Non-academic scientist | 100% | 0% | 100% |
| Total | 67% | 33% | 100% |

Some testimonies provide evidence of the positive relation between academic logic and reverse knowledge transfer, e.g. in Linguistics and Literature:

*[Contact with readers] gives me the key afterwards when it comes to explain in a lecture or in a paper some of these techniques, that there are other options [...]. [My own] professor and researcher [self] and [my own] creator and novelist [self] are lucky enough to work in the same field, it's a continuous feedback... from theory to practice and from practice to theory.*

(Academic Scientist 7)

Some degree of openness seems to be at stake in this kind of statement, let 'openness' be the broadly defined taste of scientists for engagement in usable research (Olmos-Peñuela et al. 2015), or the more concrete personality trait 'openness to experience' (McCrae and Sutin 2009). We may expect an academic scientist characterized by high levels of intellectual curiosity to be more prone to incorporate external ideas after knowledge transfer. This can also be acknowledged in the following example from Educational Science:

*The main character [of my novel] is Asperger [...]. I did some [non-academic] research for the character, although I did not have it as [an academic] research idea[10]. [My academic research is] about the effects of the introduction of new methodologies or ICTs in the classroom, e.g. the use of iPads or B-learning. In a paper, I put it as future research, [how those changes affect] that type of children with some low capabilities, like Asperger, who need routines and follow an order [...]. Then I have done some [academic] research about Asperger. Yes, the research I did for [my novel] has been useful for future research.*

(Academic Scientist 10)

Therefore, qualitative evidence suggests that, contrary to the case of direct knowledge transfer, academic logic increases other forms of knowledge transfer: indirect and reverse.

**4.4 Discussion**

This paper presents the first attempt of exploring the role that literary writing by scientists could play as a non-formal knowledge transfer mechanism. In our analyses, we have found evidences to suggest that this path deserves attention. Probably a minority of academics writes, but many writers are academics, so we believe that the above empirical findings support the formulation of propositions, which should be further explored. This may serve as a basis for more ad hoc and/or large scale empirical testing.

---

[10] Actually, let us recall, as in a previous footnote, that Academic Scientist 10 joined academia after having written the novel.

Scientific authorship of literary fiction is a prevalent phenomenon. We have examples of eminent and/or commercial scientific authors among writers. They have populated the history of literature. Even in concrete regions, we find considerable shares of scientific authors. As seen in section 4.1, they differ from non-scientific authors in their higher presence in e-books, in smaller, less potent economic regions (the Valencian Community compared to Catalonia), and in smaller publishing companies, all of which suggests a sort of marginalization of scientific authors. However, their books are typically more expensive than books of non-scientific authors, which suggests the opposite. More evidence from other regions, countries and periods is needed to reconcile these aspects.

Focusing on scientific authors, the distribution of testimonies in Table 7, plus the quotations in section 4.2.1, let us think that non-academic logic increases direct knowledge transfer via literary fiction, compared to academic logic. The higher average direct knowledge transfer of non-academic scientists compared to academic scientists (Table 9), which holds after controlling for many other variables (Table 10) confirm so. The reason is that academic scholars see this as a channel of evasion from their research activities, so lack of organizational and institutional support to literary fiction may translate into employee burnout of academics with artistic personality; on the contrary, for those academics that become fiction authors, this may constitute a source of personal development. Hence, we find that direct scientific knowledge transfer via literary fiction is feasible, but non-academic scientists tend to show higher propensity to transfer knowledge than academic scientists do.

We do find differences by field, with higher levels of alignment in Humanities, where the distinction between working and spare time for writing fiction may not be as neat as in the taxonomy presented in Table 1. The higher average value of direct knowledge transfer in Other Humanities compared to the rest of fields (Table 9), which holds after controlling for many other variables (Table 10), allow positing that belonging to the field of Humanities (other than Linguistics and Literature) increases direct knowledge transfer via literary fiction, compared to fields other than Humanities. However, the relation between writing literary fiction as a knowledge transfer strategy still needs to be better explored for Linguistics and Literature. Due to what we have named 'format-content interference', we have not been able to adopt a reliable means to measure the content alignment in such field, although our qualitative data suggest that knowledge transfer is high (see Table 8).

Our findings point out that academic scientists outperform non-academic scientists in producing indirect knowledge transfer through contacts with the cultural world and in obtaining reverse knowledge transfer for future knowledge generation. Indeed, we observe a higher frequency of interview quotes reporting indirect and reverse knowledge transfer of academic scientists, vis-à-vis non-academic scientists (Tables 11 and 12). Through the extracts from section 4.3, in the case of indirect knowledge transfer, we attribute it to influences somehow external to the scientific author, like the editors' awareness of the importance of academics, or the existence of public controversy about research topics. In the case of reverse knowledge transfer, the personality of the scientific author has a larger role, and individual openness seems to be an important prerequisite.

## 5 Conclusions

As members of publicly funded organizations, academic scientists have a central role on the creation of new knowledge in order to facilitate social progress. Lately, a key concern for policy makers is to learn about the contribution of academic scientists to social development. In this sense, contributions have been normally understood in terms of research activity and output, and not in terms of the implications of having an academic population embedded in a cultural environment. Here we build on the creative class of Florida (2005), where it is suggested that the inclusion of a certain group of individuals enhances cultural development and enrichment. However, to our knowledge, studies analysing the sociocultural relationship between cities and their universities tend to focus on specific case studies (Benneworth et al. 2010; Breznitz and Feldman 2010). These studies acknowledge sociocultural benefits, but focus on socioeconomic development and the nurturing of creative industries, and do not focus on knowledge transfer mechanisms that may trigger sociocultural outcomes in cities. We propose to analyse the role of scientists in literary fiction and, specifically, the use of this channel for knowledge transfer.

The consideration of scientific authorship of literary fiction as a knowledge transfer mechanism has led us to a theoretical contribution to existing taxonomies that distinguish formal and informal mechanisms: if scientific authorship of literary fiction leads to knowledge transfer, it would fit into a new category of non-formal mechanisms. Scientists do not use them for the sake of transfer but transfer is an important constituent, and scientists use them in their spare time rather than during office hours, as opposed to formal an informal mechanism. This concept opens two lines of inquiry. First, a conceptual one: are there any other mechanisms that fit into the category of non-formal mechanisms? Second, one with practical implications: should university managers try to internalize non-formal mechanisms and make them part of the organization?

We have also conceptualized the role of literary fiction as a vessel of knowledge transfer. Within the field of Innovation Studies, an alternative interesting approach would have been to establish an analogy between academic writers and academic entrepreneurs. We have already mentioned that the difference between academic scientists and non-scientists can influence performance same as the difference between academic and non-academic entrepreneurs (Ayoub et al. 2016). We could also ask to what extent academic authorship of literary fiction is similar to academic entrepreneurship, and hence whether it has a negative effect on scientific publication and co-publication performance (Barbieri et al. 2016). Taking the analogy a step beyond, we may elucubrate that both academic writers and entrepreneurs are seeking for further rewards beyond academy. While monetary rewards will probably not be one of academic writers' main motivations, our findings suggest some similarities with other extrinsic motivations such as the 'immense satisfaction from engaging in challenging and creative activities' (Lam 2011). In this regard, we consider that further research should seek for detailing motivations that lead scientists to engage on scientific authorship of literary fiction. Clearly, most of our interviewees are like 'serial entrepreneurs' since most of them (11 out of 14) had published repeatedly, before and/or after our target year (2015).

The benefits of academic engagement in writing literary fiction indicate the convenience of introducing this topic in the research agenda of scholars interested in knowledge transfer. Through the paper, we have shown that a related relevant problem is the burnout of academic employees with artistic personality who find lack of support by their organizations. Taken to the extreme, the ultimate problem would be that of a brain drain, i.e. whether scientific writers leave academia for literature, and hence whether academia is losing individuals with a particular skill that could be better fuelled within their organizations. Famous examples in our sample of academics leaving academia after major success include Diana Gabaldón (ex Northern Arizona University) and María Dueñas (ex University of Murcia), authors of the best sellers 'Outlander' and 'The Time in Between', respectively, both adapted for television. However, there are examples of the contrary, e.g. Umberto Eco, who has continued with his literary career and pursued it as part of his academic life. Systematic exploration of these issues would be convenient.

After our work, it would be tempting to draw conclusions about the convenience of including scientific authorship of literary fiction into the evaluation of academic merit, especially if it involves knowledge transfer. However, the research line is at too early stage to propose clear recommendations about changes in promotion incentives. Milder initiatives that would foster a creative culture among academics would be improving centralised data collection at universities about the literary activities of their faculty; and creating institutional webpages to exhibit that kind of information. There are already examples of this, e.g. in the Complutense University of Madrid[11].

This consideration leads to another potentially problematic aspect around the topic of scientific writers. The institutional implications of considering literary work within academics' scholarly works could be especially important in Humanities, as this field has been traditionally neglected within universities' priorities (Taylor et al. 2013). The promotion (within or outside of the traditional scientific evaluation system) of creative and artistic facets of scholars could enhance their presence and highlight their value within their institution. In a time when the Arts and Humanities are undervalued and considered secondary, with funding devoted to them being diminished (Deb 2017), searching and emphasizing alternative and creative venues which are already being used by researchers for knowledge transfer such as writing fiction could highlight the importance of the role of such fields.

This is just the beginning of much possible further research. We find differences between the external perception and the authors' self-perception on direct knowledge transfer of their literary work, which should be re-examined. While the external method employed may not be ideal, we consider that our approach is at least feasible and multifaceted (with quantitative and qualitative data), so we hope it has been enough to raise awareness of a prevalent phenomenon and establish propositions for future research. Application of text-mining techniques for a deeper and, to some extent, automated comparison of fiction books and research papers by the same authors would be a logical follow-up route. In addition, since we have studied the presence of scientific writers among all writers only, this line could be enriched by analysing the presence of scientific writers among all scientists, so that we could answer whether scientific writers publish as much scientific output and of the same quality as other scientists. Another open question is under what conditions we may find ambidextrous scientists that can publish, commercialise and engage in non-formal knowledge transfer simultaneously (Chang et al. 2016). Finally, we may be interested to know whether scientific writers are more read than other writers are, but readership or sales figures are not publicly available, so this should be the subject of another study.

---

[11] https://biblioteca.ucm.es/escritores/ [last access 20/10/2016].

## Acknowledgements

This research was funded by project AICO/2016/A/107 of the Valencian Regional Government. Nicolás Robinson-Garcia was supported by a Juan de la Cierva-Formación Fellowship from the Spanish Ministry of Economy and Competitiveness. We are indebted to Pablo Marín Liébana (supported by CSIC's Fellowship JAE-INT 16/00455) for his work in the database, conducting interviews and sharing ideas, and to the authors interviewed for their patience and generosity. Some of them informed about their consent to be mentioned by name: Sergio R. Alarte ('Tormentas de verano'), María Ángeles Chavarría ('Mi otro yo'), Juan Pablo Heras ('De fábula'), Xavier Minguez ('Flor de carxofa'), Lluís Miret (L'ombra del mal'), Javier Navarro ('Tableaux vivants') and Fedosy Santaella ('El dedo de David Lynch'). David Barberá-Tomás, Alejandra Boni, Elena Castro-Martínez and Richard Woolley provided invaluable feedback through informal talks, and other INGENIO colleagues during a seminar presentation. Thanks as well to attendants to the presentations of the paper at the 2016 Science and Technology Indicators Conference, the 2016 Technology Transfer Conference, the 2017 Bologna Workshop 'University-Industry Collaborations and Academic Entrepreneurship' and the 2017 Druid Conference for their participation and constructive comments, especially to our discussants Dipesh Sigdell, Carmelo Cennamo and Juan Antonio Candiani.